\documentstyle[12pt,epsf]{article}
\newcommand{\SS}{{\cal S}}

\newcommand{\GG}{{\cal G}}
\newcommand{\JJ}{{\cal J}}

\newcommand{\be}{\begin{equation}}
\newcommand{\ee}{\end{equation}}
\newcommand{\ben}{\begin{eqnarray}\displaystyle}
\newcommand{\een}{\end{eqnarray}}
\newcommand{\refb}[1]{(\ref{#1})}

\begin{document}

{}~ \hfill\vbox{\hbox{hep-th/0010190}
\hbox{CTP-MIT-3034}}\break

\vskip 3.5cm

\centerline{\large \bf  Trimming the Tachyon String Field with SU(1,1)}
\vspace*{1.5ex}

\vspace*{8.0ex}

\vspace*{3.0ex}
 \centerline{\large \rm Barton Zwiebach\footnote{
E-mail: zwiebach@mitlns.mit.edu}}

\vspace*{3.5ex}

\centerline{\large \it Center for Theoretical Physics}
\centerline{\large \it Massachusetts Institute of Technology}
\centerline{\large \it  Cambridge, MA 02139, USA}

\vspace*{5.5ex}
\medskip
\centerline {\bf Abstract}

\bigskip
A discrete symmetry of the string field tachyon
condensate noted by Hata and Shinohara is identified
as a discrete subgroup of an SU(1,1) symmetry acting
on the ghost coordinates. This symmetry, known from
early studies of free gauge invariant string field
actions, extends to off-shell interactions only for very
restricted kinds of string vertices, among them the
associative vertex of cubic string field theory. It follows
that the string field relevant for tachyon condensation
can be trimmed down to SU(1,1) singlets.

\vfill \eject
\baselineskip=17pt

Sen's conjectures on tachyon condensation and D-brane annihilation
\cite{senconj} have received striking support from calculations
\cite{KS,9912249,0002237,cubiccalc} in cubic open string field theory
\cite{WITTENBSFT}, and a proof of the energetics aspect of the
conjectures
has now been obtained \cite{exact} using
boundary string field theory \cite{boundary}. Exact solutions for
tachyon condensates in cubic string field theory are highly desirable,
for due to the connection of this string field theory to star-algebras,
homotopy algebras, non-commutative geometry and Batalin-Vilkovisky
geometry, such solutions might offer much valuable insight into
the nature of the string field.

\medskip
There has been some progress in understanding the properties
of the string field tachyon condensate. Its universality, namely
its description in terms of Virasoro generators and ghost
oscillators common to all backgrounds, was established in
\cite{9911116}. This analysis was elaborated upon in \cite{0006240},
where a calculational method tailored to this universal basis was
developed.
Intriguing suggestions for the nature of the tachyon condensate were
made
in \cite{0010034,0010060}, and an iterative approach to the construction
of the condensate was presented in \cite{0008252}. In
searching for the tachyon condensate, say, describing full
brane annihilation, one
is trying to find a time independent  spatially homogeneous string field
$T$
solving the gauge invariant  string field equation $QT + T* T=0$.
Solving such
equations requires, at least implicitly, a choice of gauge.
It may be that a simple closed form expression for $T$ does not exist
in the Siegel gauge $b_0 T=0$, and a search for a solution must
contemplate the possibility that other gauges may yield a simple
solution. So far, however, the numerical/analytic work indicates
that a solution does exist in the Siegel gauge. In fact, in an
interesting recent work, Hata and Shinohara \cite{0009105} have
verified that the tachyon condensate solving the equations
of motion following from the gauge fixed action, solves the full
gauge invariant field equations-- this can be viewed alternatively
as a confirmation that the Siegel gauge is a good nonperturbative
gauge. In this paper our considerations are mainly tailored to
the Siegel gauge, but may have a wider application.

\medskip
It was noted in Ref.~\cite{0009105} (see the second footnote
in section 4) that the three-string
vertex coupling three  open string fields in the Siegel gauge
is invariant under a discrete ${\bf Z}_4$ symmetry transformation of the
ghost and antighost oscillators:
\be
\label{transf}
b_{-n} \to  - n \, c_{-n} \,, \quad c_{-n} \to {1\over n}\, b_{-n} \, .
\ee
This, for example, was seen to imply that the modes associated to
the level-four states $(3b_{-1}c_{-3})|\Omega\rangle$ and
$(b_{-3}c_{-1})|\Omega\rangle$ must acquire identical expectation
values in the tachyon condensate. It is simple to argue that modes
odd under this discrete transformation need not acquire expectation
values in the tachyon condensate. This implies a trimming of the
string field down to ${\bf Z}_4$ singlets.

\medskip
In view of the above observations
two natural questions arise:

\begin{itemize}

\item  Is this discrete ${\bf Z}_4$ symmetry a symmetry of any
three-string vertex, or is it  a special property of the associative
vertex
of cubic string field theory ?

\item Is this discrete ${\bf Z}_4$ symmetry part of a larger
symmetry acting on the ghost coordinates that impose a larger
set of constraints on the string field ?

\end{itemize}

The purpose of this brief note is to answer these two questions.
As far as the first question is concerned, we will show that this ${\bf
Z}_4$
symmetry is not a symmetry of every three-string vertex-- althought the
associative vertex is not the only one having such symmetry. Thus, this
symmetry
is sensitive to the off-shell extension used in defining string
field theory. This is analogous to the $K_n= L_n - (-1)^n L_{-n}$
symmetries of cubic string field theory \cite{WITTENSUSFT}, which are
also
sensitive to the particular choice of string vertex.

As far as the second question is concerned we will
show that this discrete symmetry is the ${\bf Z}_4$ subgroup
of a $U(1)$ symmetry acting on the ghost coordinates and leaving
the string vertex invariant. The generator ${\cal S}_1$ of this
$U(1)$ symmetry, together with the generator ${\cal G}$ of ghost
number (also a symmetry of the string vertex) turn out to generate
the algebra of $SU(1,1)$. This algebra is precisely the same algebra
identified by Siegel and the author \cite{siegel} and used to
construct gauge invariant free string field theory. In fact, the
gauge invariant free actions used just $SU(1,1)$ singlets.
We now see that while this symmetry was not guaranteed to survive at
the interacting level, it does survive in cubic open
string field theory.  A simple argument will show that the tachyon
condensate (in fact, any nontrivial string field in the Siegel
gauge) can be restricted to the space of $SU(1,1)$ singlets.

The trimming down of the string field tachyon condensate
with $SU(1,1)$ is an interesting example of a
symmetry particular to the chosen star product that is used to simplify
the search for a solution of the string field equations.
Finding additional symmetries of this kind might simplify the search
significantly.

\bigskip
\noindent
\underbar{String vertices with ${\bf Z}_4$ symmetry}.~
Let us recall how this symmetry arises in the
cubic open string field theory. In the Siegel gauge, the string
field action reads
\be
\label{gfaction}
S \sim {1\over 2} \langle \phi|L_0 |\phi\rangle + \,{1\over 3}
  \langle \phi | \langle \phi | \langle \phi \,| \,v_3 \rangle \,,
\ee
where the gauge fixed kinetic operator $L_0$
arises from the BRST operator $Q$
\be
\label{brst}
Q = c_0 L_0 + b_0 T_+ + \widetilde Q\,,
\ee
\be
\label{brstparts}
L_0= \sum_{n=1}^\infty \Bigl( \alpha_{-n} \cdot\alpha_n
+ n ( c_{-n} b_n + b_{-n} c_n ) \Bigr)\,,
\quad T_+ = - 2 \sum_{n=1}^\infty  n \,
c_{-n} c_n \,.
\ee
In \refb{gfaction} all reference to the ghost zero modes $(b_0, c_0)$
has been eliminated, the string field is in the Siegel gauge
$b_0 |\phi\rangle =0$, where $|\phi\rangle$ is built by acting with
the negatively moded $(b_{-n}, c_{-n})$ oscillators, with $n\geq 1$
on the vacuum $|\Omega\rangle = c_1|0\rangle$, with $|0\rangle$ the
$SL(2,R)$ invariant vacuum. One may think of the dual states
$\langle \phi|$ as built with the positively moded ghost oscillators
acting on a dual vacuum $\langle \Omega|$ which satisfies
$\langle \Omega | \Omega \rangle =1$ ($\langle \Omega|\equiv \langle 0|
c_{-1}
c_0$).

The vertex coupling the three string fields is of the form
\be
\label{thevertex}
|v_3\rangle \sim  \exp\, (E_{matt}) \,\,\exp \Bigl( - \sum_{r,s = 1}^3
\sum_{n, m
=1}^\infty c_{-n}^r\,X^{rs}_{nm}\, b_{-m}^s \Bigr) \,
|\Omega\rangle_{123}\,,
\ee
where we have focused on the ghost sector. The ${\bf Z}_4$ symmetry
\refb{transf}
requires that
\be
\label{exchange}
X^{rs}_{nm} = \, {n\over m} \, X^{sr}_{mn} \,, \quad n,m \geq 1\,.
\ee
For the cubic open string field theory the ghost Neumann coefficients
satisfy
the relations \cite{gross,0009105}:
$X^{rs}_{nm} = (-)^{r+s+1} n ({\bar N}^{rs}_{nm} -
{\bar N}^{r,s+3}_{nm})$, where
\be
\label{form}
{\bar N}^{RS}_{nm}= {1\over nm} \oint_{Z_R} {dz\over 2\pi i}
\oint_{Z_S} {dw\over 2\pi i} {1\over (z-w)^2} (-)^{n(R-1) + m(S-1)}
[f(z)]^{n(-)^R} [f(w)]^{m(-)^S} \,,
\ee
for $n,m\geq 1$, and
\be
f(z) = {z (z^2 - 3)\over 3 z^2 -1} \,, \quad
Z_{R=1, 2,\cdots , 6} = \{ \sqrt{3}, {1\over\sqrt{3}} , 0,
-{1\over\sqrt{3}},
-\sqrt{3} , \infty \}\,.
\ee
The exchange property \refb{exchange} requires that
$ {\bar N}^{sr}_{mn} - {\bar N}^{s,r+3}_{mn}
={\bar N}^{rs}_{nm} - {\bar N}^{r,s+3}_{nm}$. This will hold if
${\bar N}^{rs}_{nm}={\bar N}^{sr}_{mn}$, which is essentially
manifest from \refb{form}, and if ${\bar N}^{RS}_{nm}={\bar
N}^{R+1,S+1}_{nm}$.
This last property is readily established by noting that the six points
$Z_R$ are cycled by a $SL(2,R)$ transformation $T(z)$ satisfying
$T^6 (z)=z $ and $T^3(z)= -1/z$, where:
\be
T(z) = {\sqrt{3}\, z - 1\over z + \sqrt{3}}\,, \quad
f(z) = -{z\over T^{-1}(z) T(z)} \,, \quad f(T^{-1} (z))= - {1\over f(z)}
\,.
\ee

\medskip
We now ask whether this ${\bf Z}_4$ symmetry will be a symmetry of
general
three string vertices. We will restrict the search partially
by requiring the vertices to be cyclic and symmetric\footnote{A three
open string vertex is a disk with three punctures on the boundary and three local
coordinates around them. The vertex is said to be cyclic if the
coordinates map into each other by the $SL(2,R)$ transformation
that cycles the punctures. If the disk is mapped to the upper half
$z$-plane and the punctures are conventionally located at $\{-a, 0, + a
\}$
($a$ real), the vertex is said to be symmetric if the local coordinate
around the origin goes to minus itself under the map $z\to -z$.}. For
this we found it convenient to start from a general expression for ghost Neumann
coefficients given in \cite{LPP}. These authors set up the three string vertex
as a bra where the exponential contains terms of the form
$c_n^I N^{IJ}_{nm} b_m^I$. The coefficients are then:\footnote{This is
essentially
eqn.~(4.28) of the first paper in \cite{LPP} modified because we use
the vacuum $|\Omega\rangle = c_1 |0\rangle$ rather than the
$SL(2,R)$ vacuum
$|0\rangle$. The modifications are based on the remarks around
eqn.(5.11)
of \cite{LPP}. Since \cite{LPP} treats the vertex as a bra, the relation
between $N^{IJ}_{nm}$ and $X^{IJ}_{nm}$ involves a BPZ conjugation
of the ghost bilinear, which
introduces a factor of $(-)^{n+m+1}$. This factor does not affect the
form of the exchange property \refb{exchange}, which must therefore hold
for the $N$'s.}
\be
\label{gform}
N^{IJ}_{nm} = -\oint_0 {dz\over 2\pi i} \oint_0 {dw\over 2\pi i}
z^{-n+1} \Bigl( {dh_I\over dz}\Bigr)^2 \, w^{-m-2}
\Bigl( {dh_J\over dz}\Bigr)^{-1} {1\over h_I(z)\hskip-2pt -\hskip-2pt
h_J(w)}
\prod_{K=1}^3 \Bigl( {h_J(w) - h_K(0)\over h_I(z) - h_K(0)} \Bigr)\,.
\ee
In here the $h_I$ functions define the local coordinates as maps
from canonical upper half disks to the complex upper half plane,
with $h_I(0)$ the location of the punctures. Let us fix
conventionally these
positions at $\{ - \sqrt{3}, 0, \sqrt{3}\}$ and focus on the case
when $I=J$ refers to the puncture at the origin. Denoting
by $F$ the inverse function to $h_I$ ($z = h_I(w) \to w = F(z)$),
we have
\be
\label{gsform}
N^{II}_{nm} = -\oint_0 {dz\over 2\pi i}[F(z)]^{-n+1}
\oint_0 {dw\over 2\pi i}[F(w)]^{-m-2}\, {[F'(w)]^2\over
[F'(z)]} {1\over z-w} {w (w^2-3)\over z (z^2-3)}\,.
\ee
This expression can be cast in a more symmetric way with the
use of explicit derivatives:
\be
\label{gsform}
N^{II}_{nm} = {1\over m} \oint_0 {dz\over 2\pi i}
\oint_0 {dw\over 2\pi i}[F(z)]^{-n} \Bigl\{ {d\over dw}
[F(w)]^{-m}\Bigr\}\,
{1\over z-w} {H(w)\over H(z)}\,,
\ee
where
\be
\label{introH}
H(z) = z(z^2-3) \,{F'(z)\over F(z)}\,.
\ee
Using \refb{gsform}, it is now simple to show  that the Neumann
coefficients $N^{II}_{nm}$ will have the desired ${\bf Z}_4$ symmetry
when the function $H$ satisfies
\be
\label{mcond}
{d\over dw} \Bigl(\, {1\over z-w} {H(w)\over H(z)}\, \Bigr)
={d\over dz} \Bigl(\, {1\over w-z} {H(z)\over H(w)}\,\Bigr)\,
\ee
Introducing $G(z) = H^2(z)$ the above equation yields:
\be
{1\over 2} \Bigl(  G'(z) + G'(w) \Bigr)  = {G(z) - G(w)\over z-w}\,.
\ee
This equation is solved by $G'''(z) =0$, which gives $G(z) = a^2(1+ bz +
cz^2)$,
where $a,b$ and $c$ are constants to be fixed. Back in \refb{introH} we
get:
\be
{F'(z)\over F(z)} = {a \sqrt{1+ bz + cz^2}\over z(z^2-3)}\,.
\ee
The condition that the vertex be symmetric (see footnote (2))
requires $F(-z) = - F(z)$, and this implies that $b=0$. Thus,
the condition of ${\bf Z}_4$ symmetry demands
\be
\label{gformsol}
{F'(z)\over F(z)} = {a \sqrt{1+ cz^2}\over z(z^2-3)}\,.
\ee
which gives a two parameter family of string vertices defined
by $F(z)$. While
the above expression could be integrated and the general solution
obtained, we limit ourselves to show that the Witten associative
vertex emerges as a particular solution.  Indeed, for this vertex
one has (see \cite{0006240} eqn.(2.11))
\be
\bar F(z) = \tan \,\Bigl(\, {3\over 2} \tan^{-1} (z) \,\Bigr) = {1-3z^2
- (1+
z^2)^{3/2}
\over z (z^2-3)}\,.
\ee
A straightforward computation shows that $\bar F'/\bar F$ is of the
form in \refb{gformsol} with $a=-3$ and $c=1$. This concludes our
proof that the ${\bf Z}_4$ symmetry is an off-shell type symmetry
of only very special types of string field vertices.

\bigskip
\noindent
\underbar{From ${\bf Z}_4$ to SU(1,1)}.~
The symmetry transformation in \refb{transf} can be
embedded in a continuous $U(1)$ symmetry as follows
\ben
\label{continuous}
b_{-n} (\theta) &=& b_{-n} \cos \theta - n c_{-n} \sin \theta \,,\cr
c_{-n} (\theta) &=& c_{-n} \cos \theta + {1\over n} b_{-n} \sin \theta
\,.
\een
These transformations, valid for all $n\not= 0$, coincide with the
action of the ${\bf Z}_4$ generator for $\theta=\pi/2$. Moreover,
they imply $\{ c_n(\theta), b_m(\theta) \} = \delta_{n+m}$. One readily
finds that
\be
\label{ghtrans}
b_{-n} (\theta) = e^{\theta \SS_1} \, b_{-n} e^{-\theta \SS_1}\,, \quad
c_{-n} (\theta) = e^{\theta \SS_1} \, c_{-n} e^{-\theta \SS_1}\,,
\ee
where the operator $\SS_1$ is given by
\be
\label{fgen}
\SS_1 = \sum_{n=1}^\infty \Bigl( \, {1\over n} \, b_{-n} b_n  - n c_{-n}
c_n\Bigr)\,.
\ee
We now explain why the vertex $|v_3\rangle$ is invariant under
this $U(1)$ symmetry. For this we must have
$\exp \Bigl( \theta (\SS_1^{(1)}+\SS_1^{(2)}+ \SS_1^{(3)}) \, \Bigr)
| v_3\rangle_{123} = | v_3\rangle_{123}$,
or equivalently,
\be
\label{finv}
\Bigl(\SS_1^{(1)}+\SS_1^{(2)}+ \SS_1^{(3)} \Bigr)
| v_3\rangle_{123} = 0\,.
\ee
Given eqn.~\refb{exchange},
the argument of the exponential in $|v_3\rangle_{123}$
(see \refb{thevertex})
can be written as a sum  of terms of the form
($r,s,n,m$, not summed)
\be
\label{verpart}
X^{rs}_{nm} \, c_{-n}^r b_{-m}^s + X^{sr}_{mn} \, c_{-m}^s b_{-n}^r
= {1\over m} X^{sr}_{mn} \Bigl( \, n c_{-n}^r b_{-m}^s + m c_{-m}^s
b_{-n}^r
\Bigr)\,.
\ee
It is now simple to verify that such terms commute with the
$U(1)$ generators, namely
\be
\Bigl[ \,\sum_{k=1}^3 \SS_1^{(k)}\, , \,n c_{-n}^r b_{-m}^s + m c_{-m}^s
b_{-n}^r
\Bigr]  =0 \,.
\ee
This, together with $\SS_1^{(k)} |\Omega\rangle_{123} =0$ establishes
\refb{finv}, and therefore the $U(1)$ invariance of the vertex.

Since the generator $\SS_1$ does not have definite ghost number,
it is of interest to examine its commutator with the ghost number
generator $\GG$
\be
\label{ghgen}
\GG = \sum_{n=1}^\infty \Bigl( \, c_{-n}b_n - b_{-n} c_n \, \Bigr).
\ee
Note that since the vertex $|v_3\rangle_{123}$ is built from ghost
bilinears
of zero ghost number, we have that $\GG$ is conserved:
\be
\label{ginv}
\Bigl(\GG^{(1)}+\GG^{(2)}+ \GG^{(3)} \Bigr)
| v_3\rangle_{123} = 0 \,.
\ee
The commutator mentioned above gives
\be
\label{sgen}
[ \SS_1\, , \,\GG \, ] = 2 \SS_2\,, \quad \hbox{with} \quad
\SS_2 = \sum_{n=1}^\infty \Bigl( \, {1\over n} \, b_{-n} b_n  + n c_{-n}
c_n\Bigr)\,.
\ee
One readily finds the remaining commutators:
\be
\label{rcomm}
[ \SS_2\, , \,\GG \, ] = 2 \SS_1\,,\quad
[ \SS_1\, , \,\SS_2 \, ] = -2 \GG\,.
\ee
These relations show that $\{ \SS_1, \SS_2, \GG \}$ generate
the algebra of $SU(1,1)$. These generators
are the same as those in the $SU(1,1)$ algebra of
\cite{siegel}.\footnote{Defining $X= (\SS_2 - \SS_1)/2$,
$Y=(\SS_2 + \SS_1)/2$, and $H= \GG$ we recover the conventional
definition of the isomorphic (real) Lie algebra $sl(2,R)$, with
brackets $[X,Y]= H,~ [H, X] = 2X, ~ [H,Y] = -2Y$. Note that
$T_+ = - 2X$, where $T_+$ appears as the operator multiplying
$b_0$ in the BRST operator (see \refb{brst}, \refb{brstparts}).}
Since both $\SS_1$ and $\GG$ are symmetries of the three string
vertex (\refb{finv} and \refb{ginv}) we also have
\be
\label{sinv}
\Bigl(\SS_2^{(1)}+\SS_2^{(2)}+ \SS_2^{(3)} \Bigr)
| v_3\rangle_{123} = 0 \,.
\ee
In summary, the three string vertex is fully $SU(1,1)$ invariant.

The set of Fock space states built with the action of ghost
and antighost oscillators on the vacuum $|\Omega\rangle$
can be decomposed into finite dimensional
irreducible representations of $SU(1,1)$. Note that $(nc_{-n}, b_{-n})$
transforms as a doublet. As usual, from the tensor product of two
doublets one can obtain a nontrivial singlet; this is simply
\be\label{singlets}
m b_{-n} c_{-m} +  n b_{-m} c_{-n}\,.
\ee
This expression arises from the $SU(1,1)$ invariant in \refb{verpart}
when $r=s$.

It is now simple to argue that the tachyon condensate
(or simply the string field in the Siegel gauge) can be restricted
to $SU(1,1)$ singlets. First note that the kinetic term is
$L_0$ and this operator commutes with all the $SU(1,1)$ generators.
It thus follows that the kinetic term cannot couple a non-singlet
to a singlet. Indeed consider such term $\langle s| L_0 |a\rangle$,
where $\langle s|$ denotes a singlet and $|a\rangle$ is not a singlet.
Given the structure of the representations (completely analogous
to the finite dimensional unitary representations of $SU(2)$), it
follows that there is a state $|b\rangle$ and an $SU(1,1)$ generator
$\JJ$ such that $|a\rangle = \JJ |b\rangle$. Therefore
$\langle s| L_0 |a\rangle = \langle s| L_0 \JJ|b\rangle
= \langle s|\JJ L_0 |b\rangle =0$, where the last step gives zero
because
$\JJ$ annihilates the singlet (this requires that under BPZ conjugation
$\JJ$ go into itself or minus itself, which is the case for the
$SU(1,1)$ generators). It remains
to show that the vertex cannot couple a non-singlet to two singlets.
Indeed, with analogous notation we have
\be
{}_1\langle s_1|{}_2\langle s_2| {}_3\langle a \, | v_3 \rangle
= {}_1\langle s_1|{}_2\langle s_2| {}_3\langle b \, | \JJ^{(3)}| v_3
\rangle
=- {}_1\langle s_1|{}_2\langle s_2| {}_3\langle b \, | (\JJ^{(1)}+
\JJ^{(2)}) |\,v_3 \rangle =0 \,,
\ee
where we used the conservation of $\JJ$ on the vertex, and on the
last step the $\JJ$ operators annihilate the singlets. Since
non-singlets do not have non-vanishing one point functions with
singlets, the string field can be truncated consistently to singlets,
as we wanted to show.

It is amusing to test explicitly the decoupling of non-singlets using
the available computations of the string field tachyon condensate.
The ${\bf Z}_4$ symmetry can be tested using the results indicated
in appendix A of the work of Moeller and Taylor \cite{0002237}. One
readily
checks that the  expectation values of states combine as to form the
singlets of
\refb{singlets}. At level ten the condition of $SU(1,1)$ invariance
is stronger than that of ${\bf Z}_4$ invariance. While the
level ten expectation values were not written down in their paper,
the required values, provided by the authors of \cite{0002237}
imply that the states built with four ghost oscillators appear in
the condensate as:
\ben
\label{test}
a_1 (6 b_{-1} b_{-2} c_{-3} c_{-4} + b_{-3} b_{-4} c_{-1} c_{-2} )
+ a_2 (8 b_{-1} b_{-3} c_{-2} c_{-4} + 3b_{-2} b_{-4} c_{-1} c_{-3} )\cr
+ a_3 (3 b_{-1} b_{-4} c_{-2} c_{-3} + 2b_{-2} b_{-3} c_{-1} c_{-4} ),
\een
with
\be
\label{numbers}
a_1 = 1.1472416703 \times 10^{-3}, \quad a_2 = -1.36289011 \times
10^{-6},
\quad a_3 = -1.1499674505\times 10^{-3}.
\ee
Note that each term in parenthesis in \refb{test} is ${\bf Z}_4$
invariant. Nevertheless they are not separately $SU(1,1)$ singlets. A
short
computation shows that $SU(1,1)$ invariance of the state in \refb{test}
requires
\be
a_1 - 2 a_2 + a_3 =0\,.
\ee
The values indicated in \refb{numbers} satisfy accurately this
constraint.

\medskip
The counting of $SU(1,1)$ singlets was discussed in \cite{siegel},
where using bosonization it was shown that the number of such singlets
at any level (above the vacuum $|\Omega\rangle$) is the same as the
number of states build acting on a vacuum with oscillators $a_2^\dagger,
a_3^\dagger, a_4^\dagger, \cdots$, namely, a set of bosonic oscillators
missing
the oscillator $a_1^\dagger$. The constraint of $SU(1,1)$
singlets cuts  considerably on the number of states that
can be built from ghost oscillators at each level. For example, at level
ten the
number of ghost number zero states in the ghost sector is 23, while the
number of
$SU(1,1)$ singlets is only 12. As we go higher in level the singlets
become a progressively smaller fraction of the ghost number zero states.
While this constraint should enable numerical computations
to higher levels, its  significance will be much larger if it turns
out to be an essential ingredient in the eventual construction
of an exact analytic solution for the tachyon condensate.

\bigskip
\noindent {\bf Acknowledgments}: I am grateful to N.~Moeller
and W.~Taylor for providing the level ten expectation values used
in the text. This work was
supported in part by DOE contract \#DE-FC02-94ER40818.

\end{document}